\documentstyle[amssymb,aps,multicol,graphicx]{revtex}
\begin{document}
\title{Time-temperature superposition in viscous liquids}
\twocolumn
\author{Niels Boye Olsen, Tage Christensen, and Jeppe C. Dyre}
\address{Department of Mathematics and Physics (IMFUFA), 
Roskilde University, Postbox 260, DK-4000 Roskilde, Denmark}
\date{\today}
\maketitle

\begin{abstract}
Dielectric relaxation measurements on supercooled triphenyl
phosphite show that at low temperatures time-temperature
superposition (TTS) is accurately obeyed for the primary
($\alpha$) relaxation process. Measurements on 6 other molecular
liquids close to the calorimetric glass transition indicate that
TTS is linked to an $\omega^{-1/2}$ high-frequency decay of the
$\alpha$ loss, while the loss peak width is nonuniversal.
\end{abstract}


Liquids approaching the calorimetric glass transition have very
high viscosities and very long average relaxation times compared
to, e.g., room-temperature water. The study of highly viscous
liquids is fascinating because -- independent of the chemical
nature of the intermolecular bonds -- these liquids share a
number of common features
\cite{kau48,dav53,gol64,har76,don81,bra85,ang91,nem95,%
deb96,ang97,dyr98}. The most significant common features are: i)
non-Arrhenius temperature-dependence of the average relaxation
time, and ii) non-Debye linear response functions. This paper
addresses an old empiricism relating to the latter, {\it
time-temperature superposition} (TTS) \cite{note1}.

The average relaxation time is basically the time it takes for
the liquid structure to relax to equilibrium after a small
perturbation. This quantity is easily determined from the
frequency-dependence of quantities like dielectric constant
\cite{bot78}, specific heat \cite{bir85,chr85}, shear modulus
\cite{bir87}, or bulk modulus \cite{chr94}. Measurements are
usually reported in terms of the frequency-dependence of the
response function's imaginary part, the ``loss.'' The inverse
loss peak frequency is a measure of the average relaxation time
\cite{note2}.

When temperature is lowered any loss peak moves towards lower
frequencies. This is a rather dramatic effect in the sense that
a temperature-change of just a few percent of the calorimetric
glass transition temperature $T_g$ usually changes the loss peak
frequency by more than one decade. 

Loss peaks are often plotted in log-log plots. If the shape of
the loss peak in this plot is temperature-independent the liquid
obeys TTS with respect to the response function in question
$\chi(\omega)$. Mathematically, if $T$ is temperature and
$\tau(T)$ the average relaxation time TTS is obeyed whenever
functions $N$ and $\phi$ exist such that
$\chi(\omega,T)=N(T)\phi\left[\omega\tau(T)\right]$. The factor
$N(T)$ is usually a weak function of temperature.

When TTS applies response functions are easily measured over many
decades of frequency. This is done by combining measurements done
at different temperatures. This procedure works even if only one
or two decades of frequency are directly accessible (as in, e.g.,
mechanical relaxation probed by the torsion pendulum technique or
some of the first ac specific heat measurements \cite{chr85}).
Clearly, TTS is extremely useful if it is correct. But is this
the case?

TTS has a long history \cite{note3}. For many years TTS was
assumed uncritically and used extensively. As accuracy increased
and frequency ranges widened, more and more systems were found
not obeying TTS. It turned out that TTS-violations were not just
exceptions but, in fact, quite common. These violations were
initially explained as being due to interference from an
additional minor relaxation processes, the so-called $\beta$
relaxation often found at frequencies higher than those of the
$\alpha$ relaxation. Several problems remained, however. From a
practical point of view, if TTS does not always work it cannot be
assumed {\it a priori}. Thus the option of constructing
wide-frequency master curves was no longer available. Also,
additional relaxation processes may always be postulated to
``explain'' TTS-violations, but what is to be learned from this? 

In this situation the nonpolymeric scientific community became
skeptical of TTS (for polymers TTS is still often assumed
\cite{polymer}). For several years now there has been little
interest in TTS for nonpolymeric viscous liquids close $T_g$. In
the less viscous regime at higher temperatures where
mode-coupling theory is believed to apply, this theory's
TTS-prediction has received some confirmation from both
experiment \cite{got99} and computer simulations \cite{kam98}.
From the mode-coupling perspective, breakdowns of TTS in the very
high viscosity regime correlate with the well-known breakdown of
ideal mode-coupling theory here.

We are mainly interested in the behavior of [equilibrium] highly
viscous liquids just above $T_g$. {\it If} TTS is general this
would give an important input to theory. Similarly, connecting
TTS to other features of the relaxation would also be important.
Our motivation for reinvestigating the validity of TTS close to
$T_g$ is the following. Dielectric relaxation often  exhibits a
high-frequency tail a few decades above the $\alpha$ peak, known
as the Nagel wing. The Nagel wing is temperature-dependent and
plays an important role in TTS-violations \cite{dix90,leh97}.
Although generally thought to be part of the $\alpha$ process
\cite{dix90,hof94}, it has recently been suggested that the Nagel
wing instead derives from an additional, partly hidden $\beta$
process  \cite{ols97,nga98,wag99}. A few months ago this was
confirmed by long-time annealing experiments on propylene
carbonate and glycerol by Schneider, Brand, Lunkenheimer and
Loidl \cite{sch00} (for glycerol $\beta$ relaxation was reported
at high pressures long time ago \cite{joh72}). The obvious
question now is: What happens at temperatures low enough that any
influence of $\beta$ relaxation is eliminated -- is it possible
that $\alpha$ relaxation does obey TTS after all? 

Our dielectric set-up, which is briefly described in Ref.
\cite{ols00}, covers frequencies from 1 mHz to 1 MHz. The
dielectric loss is determined with a precision better than $10^{-
4}$. The absolute temperature uncertainty is below 0.1 K with
relative variations during a frequency sweep below 1 mK
\cite{note4}.

\begin{figure}[tbp]
\begin{center}
\includegraphics[clip,width=6cm]{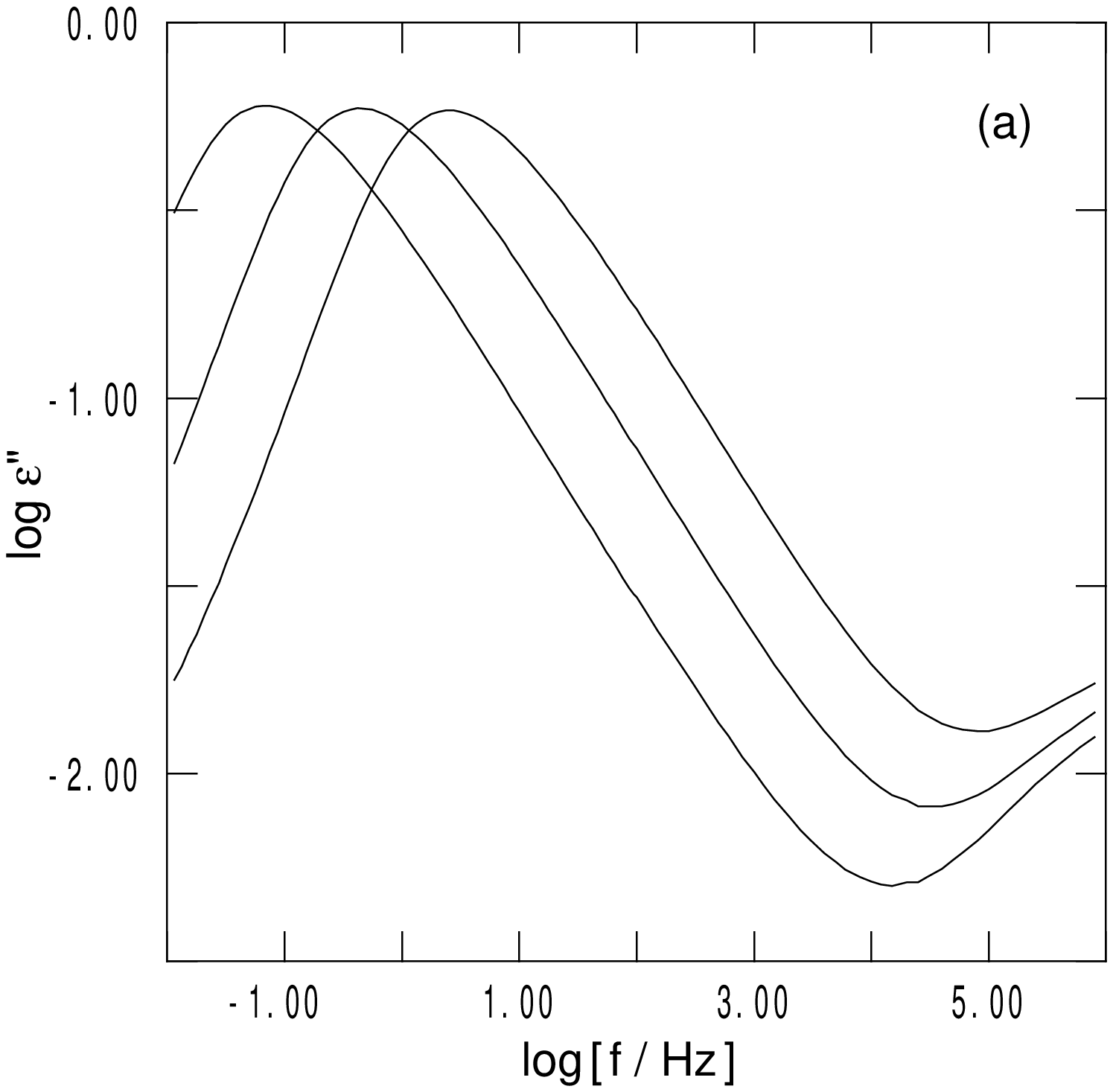}

\includegraphics[clip,width=6cm]{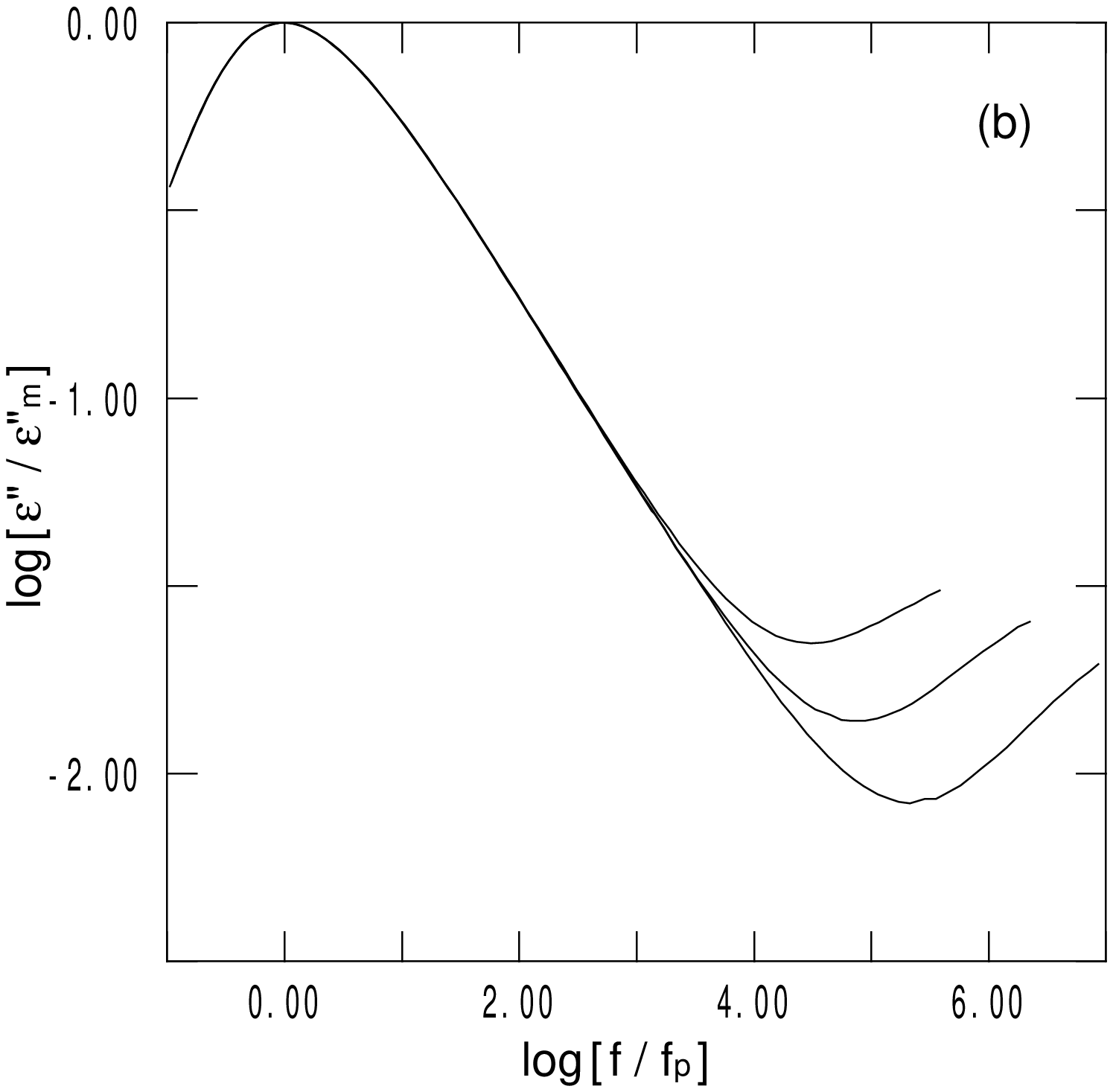}

\includegraphics[clip,width=6cm]{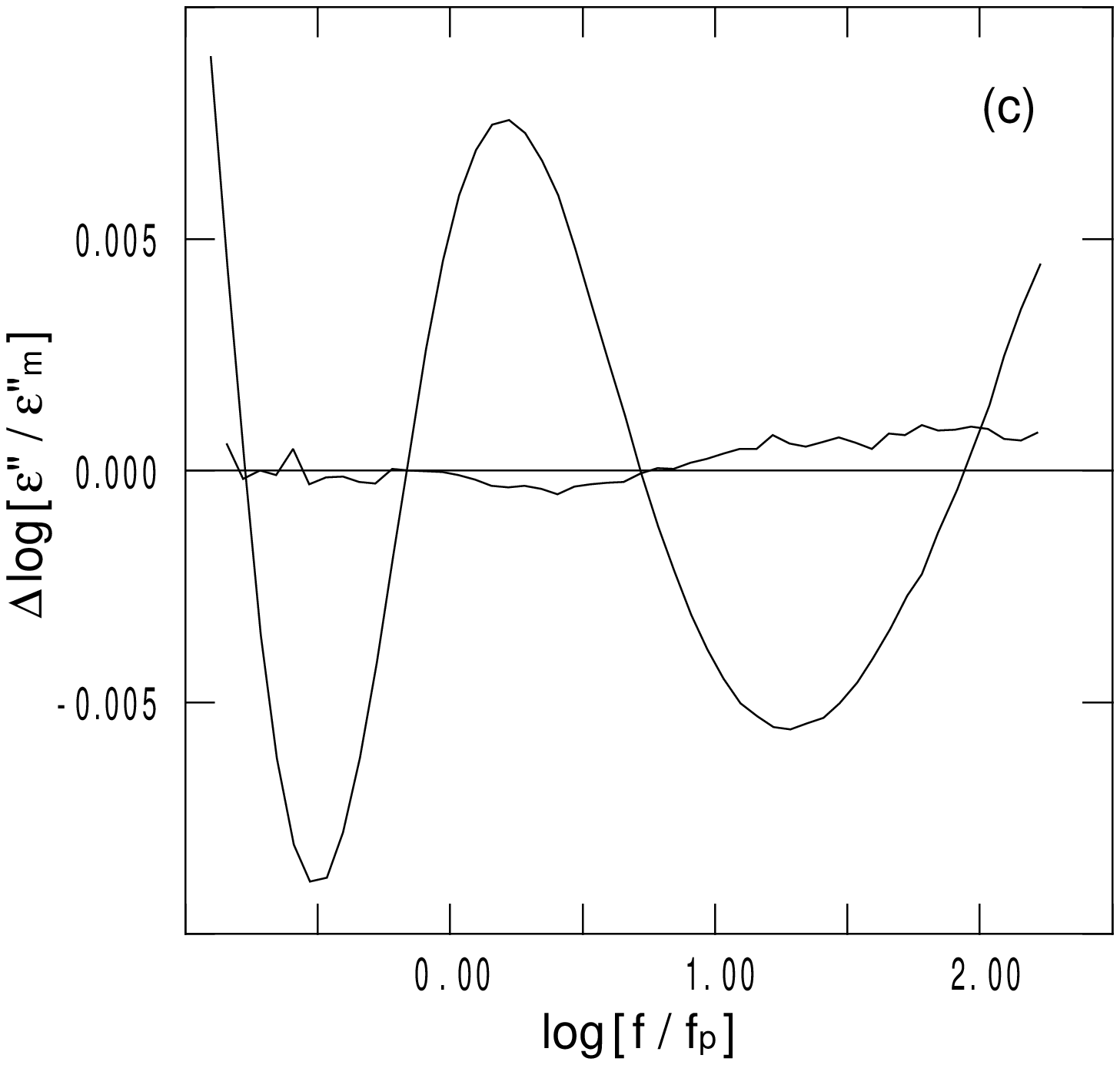}
\end{center}
\caption[jor]{Time-temperature superposition (TTS) for dielectric
relaxation of triphenyl phosphite. To avoid the ``glassy'' phase
\cite{trip_ref} the liquid was cooled fast to 210 K. (a) Log-log
plot [base 10] of the dielectric loss (imaginary part of
dielectric constant) as a function of frequency $f$ at 206.0,
208.0, and 210.0 K. As temperature is lowered the loss curve is
displaced towards lower frequencies. The figure gives raw data
(16 points per frequency decade); below 10 mHz the loss increases
due to a DC contribution [not shown]. (b) Same data scaled by the
maximum loss $\epsilon''_m$ on the y-axis and by the loss peak
frequency $f_p$ on the x-axis. TTS is obeyed at low frequencies
but violated at high frequencies where $\beta$ relaxation sets
in. (c) Detailed test of TTS. The noisy curve gives the
difference between the curves in (b) at 206.0 and 208.0 K, the
sinusoidal curve gives the difference between the 206.0 K curve
in (b) and its best Havriliak-Negami fit.} 
\label{fig1} 
\end{figure} 

Figure 1 shows the dielectric loss for triphenyl phosphite
[Aldrich, 97\%]. Figure 1a gives the raw data, Fig. 1b shows the
same data scaled to check TTS. Visually, TTS is obeyed. A closer
look is provided by Fig. 1c which compares the difference [noisy
curve] between two curves of Fig. 1b to the difference
[sinusoidal curve] between one curve and its best
Havriliak-Negami fit \cite{jon96,note5}. We conclude that TTS is
obeyed quite well. In fact, deviations are roughly as expected
from the 1 mK temperature inaccuracy.

For triphenyl phosphite $\alpha$ and $\beta$ relaxations are
unusually well separated. This is because the $\beta$ peak is at
higher frequencies than for most other viscous liquids. Two
things are to be noted from Fig. 1. First, TTS applies for the
$\alpha$ relaxation whenever temperature is low enough that
$\beta$ relaxation does not interfere. Secondly, the
high-frequency slope of the $\alpha$ peak is close to $-1/2$. How
general are these findings? Is TTS universally obeyed whenever
additional relaxations do not interfere? Is the high-frequency
$\alpha$ slope close to $-1/2$ whenever TTS is obeyed? These
questions are looked into in Fig. 2.

Figure 2a shows how $\alpha$ peak widths vary with temperature
for triphenyl phosphite ($\bullet$) and 6 other molecular
liquids. This is done \cite{dix90} by giving $w$, the logarithm
of the width at half maximum relative to a Debye process, as a
function of loss peak frequency -- the latter quantity providing
a convenient measure of temperature. When TTS is obeyed $w$ is
temperature-independent. Most liquids seem to converge to TTS at
low temperatures (the same is seen in the data of Nagel and
coworkers from 1990 \cite{dix90}), but this is not without
exceptions. In these cases (e.g., glycerol) TTS may be violated
because of interference from additional relaxations, but 
it shall be hard to prove this.

\begin{figure}[tbp]
\begin{center}
\includegraphics[clip,width=6cm]{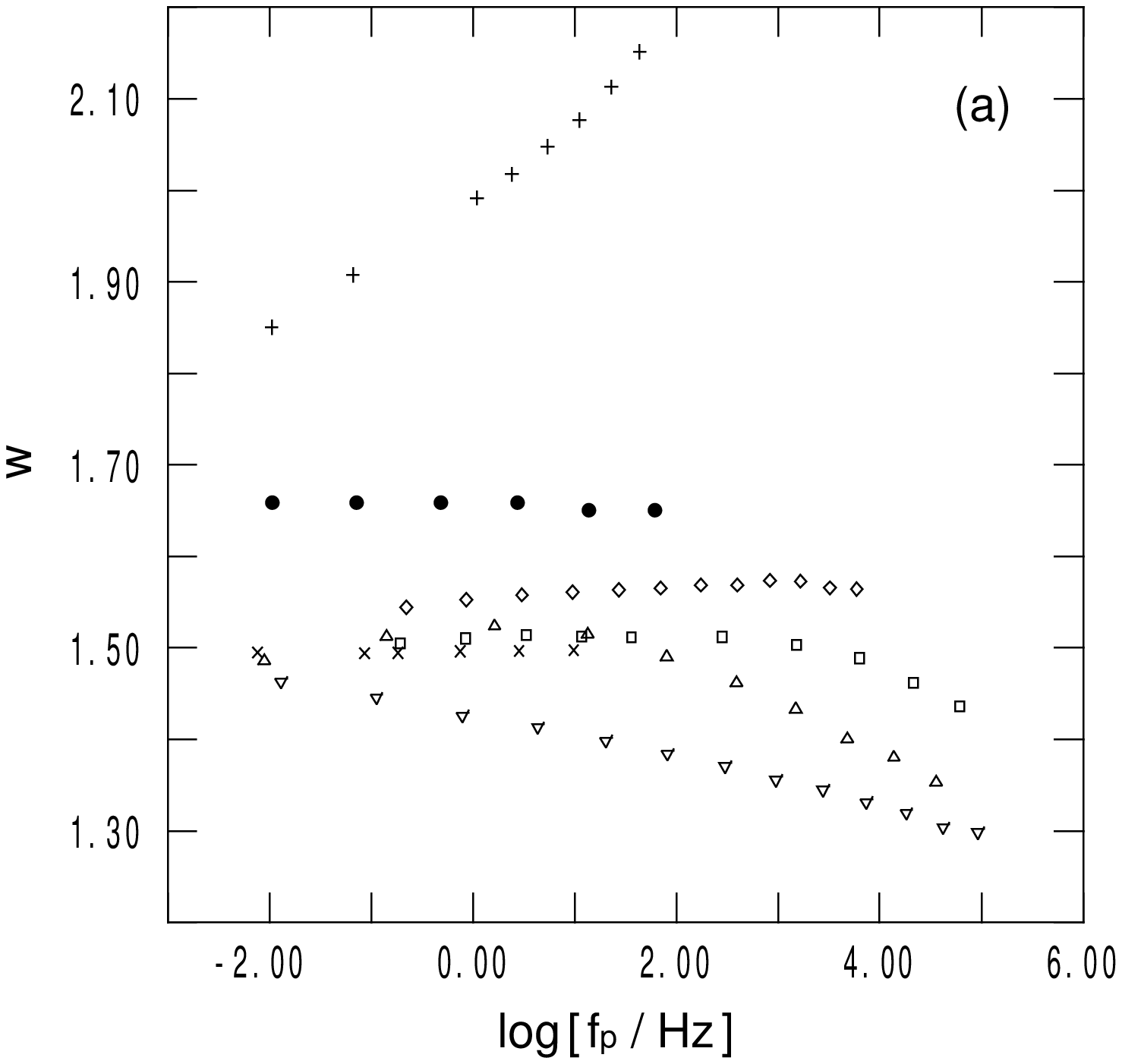}

\includegraphics[clip,width=6cm]{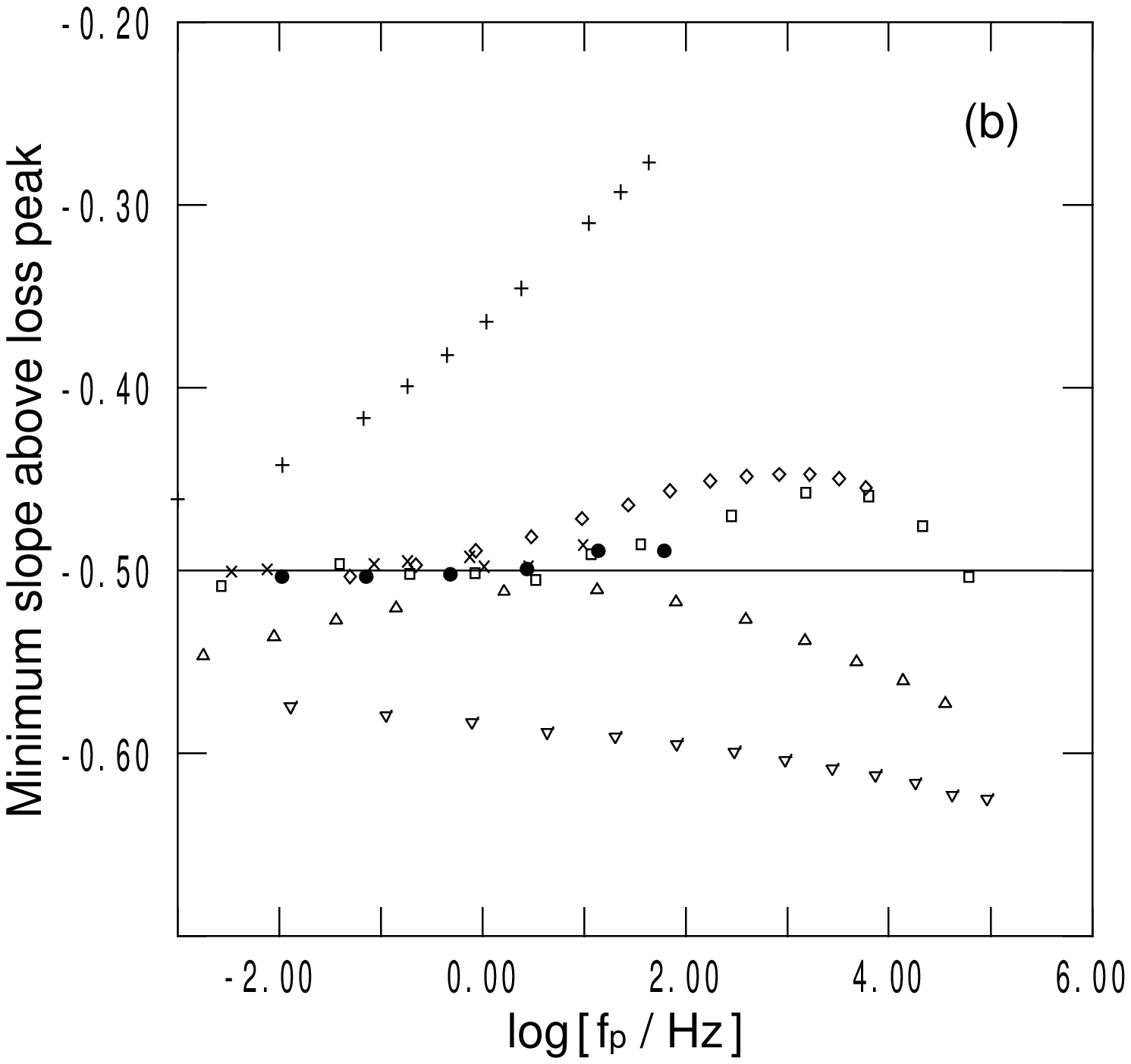}
\end{center}
\caption[lar]{Loss peak width and exponent of approximate high
frequency power-law for $\alpha$ relaxations as functions of log
[base 10] of the loss peak frequency $f_p$. The abscissa is a
convenient measure of temperature with low temperatures to the
left and high to the right. (a) Loss peak width $w$ defined by
$w=W/W_D$, where $W$ is the width at half maximum in the log-log
plot and $W_D$ is the same quantity for a Debye relaxation
process \cite{dix90}. When TTS is obeyed $w$ is
temperature-independent. This happens for some, but not all,
liquids at low temperatures. Data are shown for the following
liquids: Triphenyl phosphite [97\%] ($\bullet$), pyridine-toluene
solution \cite{ols00} ($+$), tripropylene glycol [97\%]
($\lozenge$), dipropylene glycol [techn.] ($\triangle$), dibutyl
phthalate [Sigma, unspecified purity] ($\square$), diethyl
phthalate [$>$99\%] ($\times$), glycerol [$>$99\%]
($\triangledown$). (b) Minimum slope above the loss peak of the
dielectric loss (in log-log plot) for the same 7 liquids. A
comparison with (a) shows that the minimum slope is close to $-
1/2$ whenever TTS is obeyed.}
\label{fig2}
\end{figure}

How about the $\alpha$ loss peaks for the liquids that obey TTS,
do they have anything in common? Clearly, the width is
nonuniversal. If the $\alpha$ process follows a high-frequency
power-law the exponent may be estimated by evaluating the minimum
slope above the loss peak in the log-log plot. Figure 2b shows
this minimum slope for the same 7 liquids. Comparing with Fig. 2a
we find that the minimum slope is close to $-1/2$ whenever $w$ is
virtually temperature-independent.

Figures 2a and 2b suggest that for many liquids TTS is obeyed at
sufficiently low temperatures and that, whenever this happens,
the $\alpha$ loss decays as $\omega^{-1/2}$ at high frequencies
\cite{note6}. We have indications that this picture applies also
for the frequency-dependent shear modulus\cite{beh96,upub} (which
we can measure between 1 mHz and 50 kHz \cite{chr95}), but the
shear data are more noisy than the dielectric data.

To conclude -- generalizing the triphenyl phosphite observations
-- we find that 
\begin{itemize}
\item TTS is often obeyed for the $\alpha$ relaxation at low
temperatures \cite{note9}. 
\item Whenever this happens the $\alpha$ loss decays as
$\omega^{-1/2}$ at high frequencies. 
\item The $\alpha$ loss peak width is nonuniversal.
\end{itemize}
An extension of this is the following conjecture:
\begin{itemize}
\item There is always a $\beta$ relaxation \cite{joh70}, though
it is sometimes hidden under the $\alpha$ process and only
visible as a Nagel wing.
\item Whenever temperature is so low that additional relaxations
do not interfere, the $\alpha$ relaxation obeys TTS.
\end{itemize}
To prove this beyond any reasonable doubt, however, one would
have to improve sensitivity at ultra-low frequencies and, for
some liquids, measure over many years.

To some extent the above 3+2 points reflect the opinion
prevailing in this research field 30 years ago, now generally
regarded as obsolete. Thus, Johari and Goldstein originally
conjectured that $\beta$ relaxation is ``a characteristic
property of the liquid in or near the glassy state''\cite{joh70}.
In this situation, of course, any TTS-violation gets a logical
explanation since $\alpha$ and $\beta$ relaxations have quite
different temperature-dependencies \cite{note10}. The $\omega^{-
1/2}$ high-frequency $\alpha$ peak decay was also in focus at
that time and several theories were proposed to explain this
\cite{goodoldwork}.


\begin{references}
\bibitem{note1} Other names for TTS are: Time-temperature
superposition principle, time-temperature scaling,
time-temperature reducibility, method of reduced variables, or
[for mechanical properties] thermorheological simplicity.
\bibitem{note2} Different response functions generally have
somewhat different average relaxation times (this is true even
for the bulk and shear moduli, where we find that bulk relaxation
is always slower than shear). However, in most cases response
functions have average relaxation times differing by at most 1-2
decades and their temperature-dependences are quite similar.
\bibitem{note3} Time-temperature superposition is such an
enormous subject that it is impossible to get a full view of it.
We here present the history of TTS as we recall it by reading the
literature on [non-polymeric] viscous liquids and the glass
transition during the last 30 years.
\bibitem{note4} Since the writing of Ref. \cite{ols00} the
temperature stability of our cryogenic set-up has been improved
from 5 mK to 1 mK.
\bibitem{note5} We use a Havriliak-Negami fit because it has one
more fitting parameter than other analytical approximations. If,
e.g., a stretched exponential fit is used the difference falls
outside Fig. 1c.
\bibitem{note6} At low frequencies we always find the usual
almost Debye behavior.
\bibitem{note9} $\beta$ relaxation also obeys TTS where we have
studied this in detail \cite{ols97,ols00}.
\bibitem{note10} $\beta$ relaxation is usually studied in the
glassy phase, where the loss peak frequency is Arrhenius
temperature-dependent. In the equilibrium liquid phase, however,
the $\beta$ loss peak frequency is virtually
temperature-independent \cite{ols97,ols00}.
\end{references}
\end{document}